\documentclass[conference]{IEEEtran}

\usepackage[table]{xcolor}
\usepackage{collcell}
\usepackage{hhline}
\usepackage{pgf}
\usepackage{hyperref}

\def\colorModel{hsb} 
\newcommand\ColCell[1]{
  \pgfmathparse{#1<50?1:0}  
  \ifnum\pgfmathresult=0\relax\color{white}\fi
  \pgfmathsetmacro\compA{0.08+(100-#1)/5000+.45}      
  \pgfmathsetmacro\compB{0.7+.45} 
  \pgfmathsetmacro\compC{(100-#1-7*(#1>75)+7*(#1<75))/100+.45}      
  \vspace{2pt}
  \edef\x{\noexpand\centering\noexpand\cellcolor[\colorModel]{\compA,\compB,\compC}}\x #1
  } 
\newcolumntype{E}{>{\collectcell\ColCell}m{1cm}<{\endcollectcell}}  

\usepackage{cite}
\usepackage{amsmath,amssymb,amsfonts}
\usepackage{algorithmic}
\usepackage{graphicx}
\usepackage{textcomp}
\usepackage{xcolor}
\def\BibTeX{{\rm B\kern-.05em{\sc i\kern-.025em b}\kern-.08em
    T\kern-.1667em\lower.7ex\hbox{E}\kern-.125emX}}
    
\usepackage{fancyhdr}
\begin{document}

\title{Tweet Moodifier: Towards giving emotional awareness to Twitter users}

\author{\IEEEauthorblockN{Bel\'en Sald\'ias F.}
\IEEEauthorblockA{\textit{MIT Media Lab} \\
\textit{Massachusetts Institute of Technology}\\
Cambridge, USA \\
belen@mit.edu}
\and
\IEEEauthorblockN{Rosalind W. Picard}
\IEEEauthorblockA{\textit{MIT Media Lab} \\
\textit{Massachusetts Institute of Technology}\\
Cambridge, USA \\
picard@mit.edu}
}

\maketitle
\thispagestyle{fancy}

\begin{abstract}
 Emotional contagion in online social networks has been of great interest over the past years. Previous studies have focused mainly on finding evidence of affect contagion in homophilic atmospheres. However, these studies have overlooked users' awareness of the sentiments they share and consume online. In this paper, we present an experiment with Twitter users that aims to help them better understand which emotions they experience on this social network. We introduce Tweet Moodifier (T-Moodifier), a Google Chrome extension that enables Twitter users to filter and make explicit (through colored visual marks) the emotional content in their News Feed. We compare behavioral changes between 55 participants and 5089 of their public ``friends.'' The comparison period spans from two weeks before installing T-Moodifier to one week thereafter. The results suggest that the use of T-Moodifier might help Twitter users increase their emotional awareness: T-Moodifier users who had access to emotional statistics about their posts produced a significantly higher percentage of neutral content. This behavioral change suggests that people could behave differently while using real-time mechanisms that increase their affect reflection. Also, post-experience, those who completed both pre- and post-surveys could assert more confidently the main emotions they shared and perceived on Twitter. This shows T-Moodifier's potential to effectively make users reflect on their News Feed.
\end{abstract}


\section{Introduction}
\label{sec:instroduction}
Social media users are constantly creating content that connects them to others, but are users aware of the emotional influence that social media has on their moods or lives? While consuming and sharing information online is advantageous to connecting people, it may pose a risk to ignore the emotions that we (consciously or unconsciously) spread out to hundreds or thousands of people with a single post.

Over the past few years, emotional contagion through social media has been of great interest. Kramer et al.~\cite{kramer2014experimental} argue, ``Emotional states can be transferred to others via emotional contagion, leading people to experience the same emotions \textit{without their awareness}. Emotional contagion is well established in laboratory experiments, with people transferring positive and negative emotions to others.'' In addition, several studies, e.g., \cite{kramer2014experimental, fan2014anger, coviello2014detecting, ferrara2015measuring}, have found evidence of affect contagion in homophilic atmospheres, where people tend to connect with others who are socially similar to themselves. Furthermore, Tromholt~\cite{tromholt2016facebook} and Hunt et al.~\cite{hunt2018no} show that the content that we consume on platforms like Twitter or Facebook affects not only the emotions that we express on these platforms but also our general wellbeing{\color{black}, for example, through consuming toxic content or experiencing cyberbullying attacks~\cite{wulczyn2017ex, schmidt2017survey, rosa2019automatic}.}

\begin{figure}
    \centering
    \includegraphics[width=.62\linewidth]{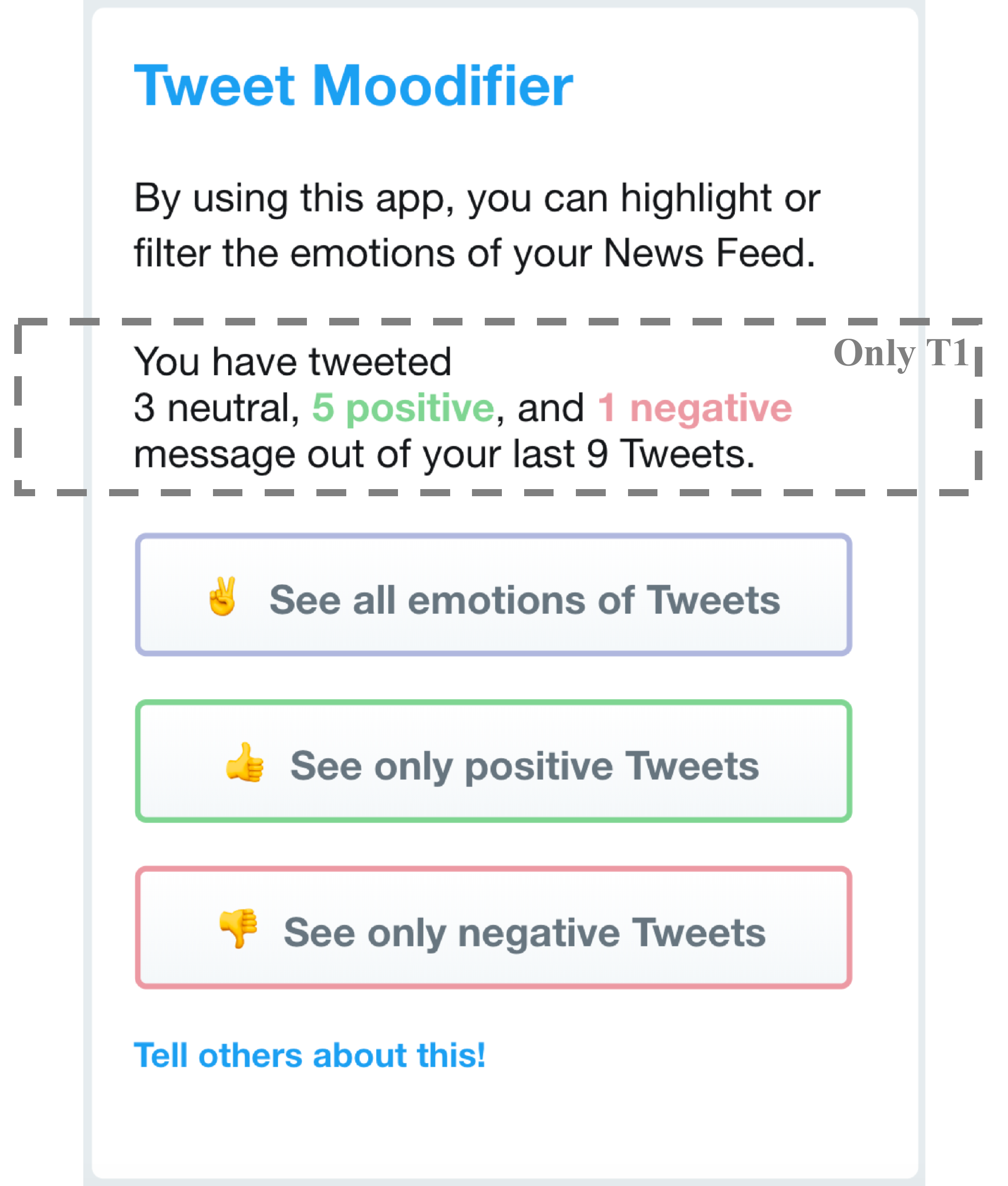}
    \caption{T-Moodifier options. Framed statistics are only present in T1.}
    \label{fig:basecard}
\vspace{-1em}
\end{figure}

However, all of these insightful studies fail to address social media users' emotional awareness, {\color{black}i.e., users' ability to identify and differentiate emotions~\cite{boden2011you, huang2013distinguishing}}, of the content that they consume and share. {\color{black} On the one hand, we can leverage emotional awareness to help users improve the regulation of their emotions on social media~\cite{erisman2010preliminary, hill2012mindfulness} by assisting them with emotional differentiation~\cite{hill2012mindfulness, barrett2001knowing}. On the other hand, it has been shown that users tend to change their behavior on social media: They aim to deliver a positive impression of themselves that they want others to perceive \cite{ahn2013show, zhao2008identity, doi:10.1111/j.1468-2958.2007.00312.x}, self-censor their posts if they cannot filter their audience~\cite{sleeper2013post}, and try not to involve themselves in conflicts that may give others an overall bad impression~\cite{rainie2012tone}. In this study, we couple these users' intentions with a new tool that aims to help increase emotional awareness. We aim to understand to what extent Twitter users are aware of or reflect on the impact caused by the emotional content that they consume and create. In addition, we examine how users' sharing patterns change after they use this new tool.}

We introduce, build, deploy, and evaluate Tweet Moodifier (T-Moodifier), a Google Chrome extension that enables Twitter users to explore the emotional sentiment of posts in their News Feed (see figure \ref{fig:basecard}). The extension is powered by a machine learning algorithm that classifies Tweets into three different sentiment categories: \textit{positive} posts, which tend to use happy or surprising language; \textit{negative} posts, which tend to use sad, angry, or disgusting language; and posts without strong emotional language, which are classified as \textit{neutral}. Using T-Moodifier, participants have access to three new views (which can be regarded as different perspectives {\color{black}that can help users with emotional differentiation~\cite{barrett2001knowing}}) of their Twitter feed: (1) highlight with colors the emotional valence of each post in their News Feed, (2) filter out all negative and neutral content to keep only positive posts, and (3) filter out all positive and neutral content to keep only negative posts. Users also have the opportunity to click an emoji (positive, neutral, or negative) below each Tweet if they wish to indicate a different label for the automatically-labeled valence. Our hypotheses are that by making users aware of the emotions that surround them online, in this case through Tweets, users could:

\begin{enumerate}
    \item Change their emotional sharing patterns.
    \item Reflect more on, and consequently be more aware of, what they post before doing so.
    \item Feel better or become mentally healthier.
\end{enumerate}

From the studies mentioned above \cite{coviello2014detecting,ferrara2015measuring}, one may conclude that showing positive content to people would make them happier. However, some researchers in the mental health community \cite{hunt2018no,primack2017social} propose that consuming only positive emotions can lead people to experience a negative feeling of isolation. Since it is not our primary goal to measure emotional contagion, T-Moodifier does not suggest that participants should spend more time in one T-Moodifier view than in another. {\color{black}Nevertheless, to prevent users from forgetting that they have the negative emotions view activated, if a user spends more than 15 minutes there, T-Moodifier shows a pop-up that blinks until the user restores his or her original feed (this pop-up was tested in a 10-person pilot without complaints, so we decided to keep it).} To the best of our knowledge, T-Moodifier is one of the first attempts to try to enhance and understand social media users' awareness of how Twitter affects them.

Results presented in this paper are based on 55 users who used T-Moodifier for at least 7 days each, for a total of 21 days studied per user (from two weeks before installing T-Moodifier to one week thereafter). Our findings suggest that the use of T-Moodifier increases users' emotional awareness. In addition, participants tend to neutralize and reflect more on their content when they have access to emotional statistics. This behavioral change suggests potential benefits of creating real-time mechanisms that increase social media users' awareness. Finally, participants who completed both pre- and post-experience surveys asserted more accurately and confidently the primary emotions they shared and perceived on Twitter.

The rest of this paper is organized as follows: In section~\ref{sec:protocol}, we describe the study protocol. In section~\ref{sec:classifier}, we explain the classifier in charge of giving emotional labels to users' Tweets. In section~\ref{sef:ppolicy}, we disclose our privacy policy. Finally, in sections~\ref{sec:results},~\ref{sec:conclusion}, and~\ref{sec:discussion} we present our results, conclusions, and discussion plus future work, respectively.

\section{Study protocol}
\label{sec:protocol}
First of all, we want to highlight that participants have full control over the T-Moodifier experience: a) They choose to use the plugin (or not), b) They have can stop using it at any time, and c) They are informed (and asked to explicitly accept the Privacy Policy; more details in section \ref{sef:ppolicy}) about which kinds of data will be collected on their usage and sharing patterns. Furthermore, we only use publicly available Twitter data.

Twitter users are invited to use T-Moodifier through its website\footnote{\url{https://tweetmoodifier.media.mit.edu}}, {\color{black}which, for this study, was sporadically promoted by the MIT Media Lab's social media accounts for two weeks\footnote{ $\sim$ 450K Twitter followers.}}. Upon installing the extension (participants are able to uninstall it at any time), they see a description of the tool's purpose and are prompted to answer a pre-experience survey and continue only if they accept our privacy policy and usage conditions.

Next, participants are randomly assigned to one of two different treatment groups, who are also compared to a control group based on public data. Participants are shown a set of options (see figure \ref{fig:basecard}) below their profile picture in the Twitter homepage, that allows them to:

\begin{itemize}
    \item Treatment 1 (\texttt{T1}): see valences of Tweets in their News Feed, filter the emotions of the Tweets that they consume, and see personal statistics revealing the valence of what they tweeted.
    \item Treatment 2 (\texttt{T2}): same as \texttt{T1} but NOT seeing any statistics about Tweets they generated.
    \item Control: A random sample of ``friends,'' defined by Twitter as people our participants follow, who do not install the plugin. {\color{black}This group could include influencer/popular accounts whenever participants followed them.}
\end{itemize}

{\color{black}We define the control group assuming that it might reflect a homophilic atmosphere for participants. Then, if participants have been in synchrony with their homophilic group until T-Moodifier is introduced, and only participants change their behavior during treatment, we can say that T-Moodifier can account for these behavioral changes and not their homophilic atmosphere.}

When someone activates any of the T-Moodifier views, sentiment classification is obtained using the Tweets emotion classifier described in section \ref{sec:classifier}. In addition, T-Moodifier activates a visual mark as a reminder of the view they have enabled; T-Moodifier displays green borders in positive Tweets, red borders in negative Tweets, while neutral Tweets remain without a colored mark. Finally, after a seven-day experience, participants are taken to a post-survey. They are reminded of access to mental health resources {\color{black}(also available in our Privacy Policy \ref{sef:ppolicy})} each time they activate a T-Moodifier view {\color{black}(see Appendix for visual details)}.

While participants were told that T-Moodifier aimed to help them better understand which emotions they consume online, we also analyzed how they behaved while using T-Moodifier. We hypothesized that, since participants in \texttt{T1} had access to personal statistics about their publicly observable behavior on Twitter, they would be most likely to show a change in behavior regarding the emotional distribution of the Tweets they create. This study was developed under the Massachusetts Institute of Technology IRB Protocol \#1810563376.

\section{Emotions classifier}
\label{sec:classifier}

Aiming to deliver an efficient application, based on predictions of an affect sentiment classifier that is well known, {\color{black}relatively easy to interpret compared to most recent techniques}, and trained for Tweets (see Giachanou and Crestani's work for a survey~\cite{giachanou2016like}), we used the model by Go et al.~\cite{go2009twitter}. Their model uses Twitter APIs to generate its training and validation datasets. The model analyzes Tweets' text, but first, authors strip out all emojis in the Tweets, to later try to predict these emojis as emotions (positive, neutral, or negative) labels as a way of distant supervision. There are several emojis that can be classified as positive (e.g. :), :-), and :D), and negative (e.g. :(, :'(, and :@). The full list of emojis can be found in Go. et al.'s paper~\cite{go2009twitter}. Thus, their model uses self-labeled ``non-verbal'' emojis as its ground truth for sentiment labeling. We used this pre-trained classifier through its available API at \url{http://help.sentiment140.com/api}.

As with most machine learning systems, this classifier can make mistakes. Its authors reported accuracy score levels above 80\%. Unintended ``accidents'' can occur and be potentially harmful for people~\cite{amodei2016concrete}. Hence, we address this issue by allowing participants to rectify the emotional classification of what they observe (Tweets are relabeled only for them and not for all T-Moodifier users). {\color{black}As we collect users' re-labeled data, we could retrain the base model to provide more accurate or personalized emotion labels. However, the results reported here are under the assumption that by rectifying mislabeled Tweets, users can reflect on more accurate valences of their Tweets; hence, we do not retrain the base model.}

\section{Privacy Policy}
\label{sef:ppolicy}
{\color{black}We acknowledge that this line of research requires critical thinking about how to exploit and interpret users' data~\cite{panger2016reassessing}.} For Twitter experiments, some users (private users) may have decided to make their Tweets only visible to their Twitter followers. Hence, even though Twitter's Privacy Policy would allow us, T-Moodifier does not read or evaluate emotions of protected Tweets when analyzing participants' News Feeds. {\color{black}We make our privacy policy fully available to the community and users at \url{https://tweetmoodifier.media.mit.edu/privacy-policy}.}

\section{Results}
\label{sec:results}

We present results for the hypotheses stated in Section~\ref{sec:instroduction}. This analysis is based on 55 users; 20 female and 33 male participants reported their gender; \texttt{T1}: 24, \texttt{T2}: 28, and protected: 3{\color{black}, who are only included in the survey analysis}. They all used T-Moodifier for at least 7 days days, yielding 385 of T-Moodifier use. For the control group, we sample up to 100 friends of each participant, reaching 5089 unique users. {Data are collected using Twitter's Python API (TweePy\footnote{\url{http://tweepy.org}}). \color{black}A pre-experience survey also captured the following information:}
\begin{itemize}
    \item Eighty-two percent of users were aged 18 to 34 years.
    \item Seventy-five percent of participants
reported that they used Twitter at least once a day, whereas
62\% of all participants stated that they used it several times a day.
    \item What option best describes what you use Twitter for?: The most popular answer (with a frequency of 51\%) among 10 options was that participants used Twitter to keep up with or share the news in general.
\end{itemize}

{\color{black}We use the t-test to compare between the treatments and control group and the paired t-test to compare behavioral changes and pre- versus post-survey responses. We use $p = 0.05$ as the cutoff for significance.}

\subsection{Sharing patterns}

We analyzed the proportion of positive, negative, and neutral Tweets posted by each member of \texttt{T1} and \texttt{T2}; {\color{black}this analysis also includes what users wrote in Tweets that retweeted others}. We compared how the average distribution, for public users, changed from before introducing T-Moodifier to one week thereafter. Table \ref{table:sharing-patterns-control-group} shows that the control group distribution of the emotions they share remains steady from two weeks prior to T-Moodifier through the week of treatment, with no statistically significant changes.

The number of Tweets pre/post experience did not change significantly, with an average of 11 and 15 Tweets per user for treatments and the control group, respectively. We define, 
\begin{itemize}
    \item $W_{-2}$: the second week before treatments.
    \item $W_{-1}$: the week before treatments.
    \item $W_{0}$: the week of treatments.
    \item \texttt{T}: All T-Moodifier participants.
\vspace{-.25em}
\end{itemize}

\begin{table}[h!]
\centering
\caption{Emotional content of \textbf{control group's} Tweets in \%.}
\newcommand\items{3}
\arrayrulecolor{white}
\begin{tabular}{l*{\items}{|E}|}
\multicolumn{1}{l}{} &
\multicolumn{1}{c}{$W_{-2}$} & 
\multicolumn{1}{c}{$W_{-1}$} & 
\multicolumn{1}{c}{$W_{0}$} \\ \hhline{~*\items{|-}|}\hhline{~*\items{|-}|}
Positive  & 29.3   & 30.3  & 28.2   \\ \hhline{~*\items{|-}|}
Neutral   & 64.9   & 64.1  & 65.9   \\ \hhline{~*\items{|-}|}
Negative  & 5.8   & 5.7  & 5.9   \\ \hhline{~*\items{|-}|}
\end{tabular}
\label{table:sharing-patterns-control-group}
\vspace{-1.5em}
\end{table}

\begin{table}[h!]
\centering
\caption{Emotional content of \textbf{T-Moodifier users'} Tweets in \%.}
\newcommand\items{4}
\arrayrulecolor{white}
\begin{tabular}{l*{\items}{|E}|}
\multicolumn{1}{l}{} &
\multicolumn{1}{c}{$W_{-2}$, \texttt{T}} & 
\multicolumn{1}{c}{$W_{-1}$, \texttt{T}} & 
\multicolumn{1}{c}{$W_{-1}$, \texttt{T1}} & 
\multicolumn{1}{c}{$W_{0}$, \texttt{T1}} \\ 
Positive & 26.8 & 28.7 & 34.7 & 16.9 \\ \hhline{~*\items{|-}|}
Neutral & 68.2 & 66.3 & 62.8 & 79.7 \\ \hhline{~*\items{|-}|}
Negative & 5.0 & 5.0 & 2.5 & 3.4 \\ \hhline{~*\items{|-}|}
\end{tabular}
\label{table:sharing-patterns}
\vspace{-1.5em}
\end{table}

\begin{table}[h!]
\hspace{1.85em}
\arrayrulecolor{white}
\begin{tabular}{p{3.9cm}*{2}{|E}|}
\multicolumn{1}{l}{} &
\multicolumn{1}{c}{$W_{-1}$, \texttt{T2}} & 
\multicolumn{1}{c}{$W_{0}$, \texttt{T2}} \\ 
Positive & 27.8 & 31.9 \\ \hhline{~*2{|-}|}
Neutral  & 69.4 & 64.3 \\ \hhline{~*2{|-}|}
Negative & 2.8 & 3.8 \\
\end{tabular}
\end{table}

As seen in Table \ref{table:sharing-patterns}, the hypothesis that participants in \texttt{T1} would change their fraction of positive, neutral, and negative Tweets while using T-Moodifier ($W_{0}$, \texttt{T1}) was confirmed. In particular, they produced mainly neutral
content. This Table also illustrates that prior to introducing T-Moodifier ($W_{-1}$ and $W_{-2}$), participants (\texttt{T}) behaved similarly to the control group.

On the one hand, from Tables \ref{table:sharing-patterns} and \ref{table:stats} we observe significant differences for T-Moodifier users under Treatment 1 (\texttt{T1}), and since there is no evidence to claim that \texttt{T1} participants behave differently from their friends prior to using T-Moodifier, we can use their friends as a control group. Furthermore, we see that while \texttt{T1} participants use T-Moodifier they behave significantly different from themselves prior to T-Moodifier. They also behave significantly different from both the control and \texttt{T2} groups during the treatment week. On the other hand, we found that participants in \texttt{T2} did not significantly change their behavior as hypothesized. The only difference between treatments is that participants in \texttt{T1} had access to personal statistics about their posts' valences. 

\vspace{-.5em}
\begin{table}[h!]
\centering
\caption{P-values for differences between groups in tables \ref{table:sharing-patterns-control-group} and \ref{table:sharing-patterns}}
\begin{tabular}{|llcc}
\multicolumn{1}{l}{} & \multicolumn{1}{l}{} & \multicolumn{1}{c}{Positive} & \multicolumn{1}{c}{Neutral}\\ \hline\hline 
$W_{-1}$ & 	$W_{-1}$, \texttt{T1} 	& $0.4122$   &  $0.8078$\\
$W_{0}$ & 	$W_{0}$, \texttt{T1} 	& $\textbf{0.0370}$     &   $\textbf{0.0115}$    \\ \hline
$W_{-1}$, \texttt{T1} & $W_{0}$, \texttt{T1}   &  $\textbf{0.0305}$  &  $\textbf{0.0351} $  \\
$W_{0}$, \texttt{T2} & 	$W_{0}$, \texttt{T1}   &  $\textbf{0.0459}$ & 	$\textbf{0.0482}$   \\\hline
\end{tabular}
\label{table:stats}
\end{table}

Hence, the results suggest that T-Moodifier, especially with personal feedback about the valence of what a user is posting, can potentially affect the affective valence of what that user tends to share on social media. {\color{black}These results resonate with Grosser's~\cite{grosser2014metrics} argument about Facebook users being driven by social metrics.}

Note that shifting toward a highly neutral distribution is not necessarily an expected change. Social media are powerful platforms for discussing social injustice issues that may make people experience negative emotions. Introducing T-Moodifier is not an attempt to calm down those essential human needs. Feeling negative or sharing negative information is sometimes very healthy, as is feeling positive and sharing positive information. 

Table \ref{table:stats} shows the results of testing for (and confirming) the statistical significance of the hypothesized difference between $W_{0}$ and ($W_{0}$, \texttt{T1}). We also see confirmed a significant difference between ($W_{-1}$, \texttt{T1}) and ($W_{0}$, \texttt{T1}) as well as between ($W_{0}$, \texttt{T2}) and ($W_{0}$, \texttt{T1}). We present a summary of the standard deviations for these interesting sharing patterns changes in Table \ref{table:c-intervals}.

\vspace{-.75em}
\begin{table}[h!]
\centering
\caption{Standard deviations in \% of results in Tables \ref{table:sharing-patterns-control-group} and \ref{table:sharing-patterns}}
\begin{tabular}{lccccc}
\multicolumn{1}{c}{} & 
\multicolumn{1}{c}{$W_{-1}$} &
\multicolumn{1}{c}{$W_{-1}$, \texttt{T1}} & 
\multicolumn{1}{c}{$W_{0}$} & 
\multicolumn{1}{c}{$W_{0}$, \texttt{T1}} & 
\multicolumn{1}{c}{$W_{0}$, \texttt{T2}} \\ 
Positive & 26.2 & 30.7 & 26.5 & 19.4 & 31.5 \\
Neutral  & 26.1 & 30.8 & 26.7 & 21.3 & 31.1 \\
Negative  & 10.3 & 6.6 & 11.4 & 5.7 & 1.1 \\
\end{tabular}
\label{table:c-intervals}
\end{table}

Finally, we looked at the participants' engagement with T-Moodifier features. Participants used T-Moodifier views 25.4\% of the times they used Twitter (77\% of the time they preferred to highlight all three valences; 12\% only positive valences, and 11\% only negative valences). While using the T-Moodifier views, participants displayed 37.5 Tweets on a daily average (std: 57.5). Regarding mislabeling, 35\% of users relabeled 3\% of the Tweets displayed. We did not find tendencies toward specific mislabeled emotions.

\subsection{Self-reported awareness}
\label{sec:self-reported-aw}
How was awareness of mood assessed and was use of T-Moodifier associated with any changes in self-reported awareness of how Twitter impacts mood? We presented the participants with a pre- and post-questionnaire, each time asking them the following questions, whose responses are discrete numbers from 1 to 100 according to the stated limits (e.g. ``never - always,'' ``insignificant - enormous'').

\begin{enumerate}
    \item[(1)] Twitter influences my mood (never - always).
    \item[(2)] My connections on Twitter influence the emotions that I experience (never - always).
    \item[(3)] Twitter has (insignificant - enormous) influence on the mood of others.
    \item[(4)] I influence the emotions that others experience on Twitter (never - always).
    \item[(5)] I am aware of (or reflect on) the emotions expressed in my Tweets before I post them (never - always).
    \item[(6)] My confidence in the response to the question: ``Most of my Tweets are emotionally: negative or neutral or positive'' is (extremely weak - extremely strong).
    \item[(7)] My confidence in the response to the question: ``Most of my friends' Tweets are emotionally: negative or neutral or positive'' is (extremely weak - extremely strong).
\end{enumerate}

\vspace{-.5em}
\begin{table}[h!]
\centering
\arrayrulecolor{black}
\caption{Pre-survey --- Twitter emotional influence questions.}
\begin{tabular}{cccccccc}
& \multicolumn{1}{c}{(1)} & \multicolumn{1}{c}{(2)} & \multicolumn{1}{c}{(3)} & \multicolumn{1}{c}{(4)}  & \multicolumn{1}{c}{(5)} & \multicolumn{1}{c}{(6)} & \multicolumn{1}{c}{(7)} \\ \hline
\textbf{Pre}, \texttt{T} &   &   &   &   &   &   &   \\
Mean & 44.5 & 46.3  &  52.9 &  \textbf{21.6} &  61.2 &  58.7 &  50.2 \\
Std & 28.1  & 24.1  &  28.5 &  22.1 &  30.9 &  31.6 &  24.3 \\\hline\hline
\multicolumn{7}{l}{Note that row (\textbf{Pre}, \texttt{T}) are results for all 55 participant.}
\end{tabular}
\label{table:surveys-1}
\end{table}

\subsubsection*{\textbf{Pre-survey results}, \texttt{T}}

In Table \ref{table:surveys-1}, from Questions (1), (2), and (3) we see that T-Moodifier users recognize that Twitter can influence emotions to some extent but only occasionally. However, (see Question (4)) they strongly believe that their Tweets or interactions on Twitter \textit{almost never} influence the emotions of others. 

From Question (5), we see that the participants declare that they are somewhat aware of the emotions of what they share. However, they declare in Question (6) and (7) that they are not so sure about the emotional content they spread through, or consume from, social media. These last three questions show that users may recognize that what they consume or create has emotional content, and they may reflect on it somewhat, but they are not very confident as to whether the overall tone of what they are sending and receiving is more negative or more positive.

\begin{figure}[h!]
    \centering
    \includegraphics[width=.6\linewidth]{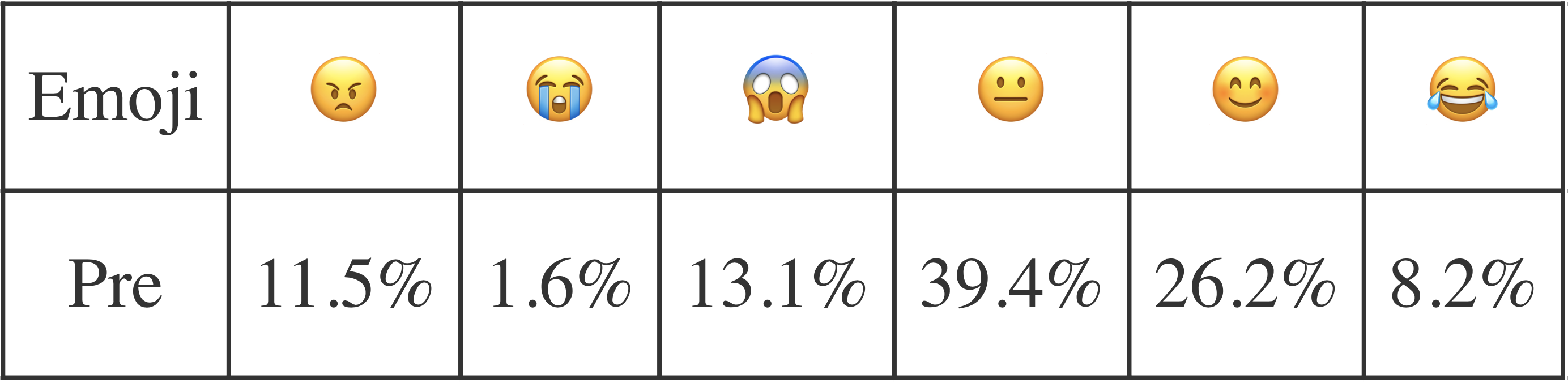}
    \caption{Pre-questionnaire responses to: ``The emoji that best describes how scrolling through Twitter makes you feel is \dots'''.}
    \label{fig:emojis}
\end{figure}

We also asked them to point to the emoji that best describes how scrolling through Twitter makes them feel; the results are shown in Figure~\ref{fig:emojis}. The large bias toward neutral might be associated with the fact that 51\% of the users claimed that they use Twitter mostly to keep up with or share the news in general (which could be thought as informative content) or it could also be associated with recognizing that they have a lot of both positive and negative emotions to things they see on Twitter, but they were forced to select only one emoji.

\subsubsection*{\textbf{Post-survey results, *}} In this section, we present results regarding only those participants who completed both pre-questionnaire (\textbf{Pre}, *) and post-questionnaire (\textbf{Post}, *) responses. They constituted 28 people (\texttt{T1}: 16, \texttt{T2}: 12).

\vspace{-.5em}
\begin{table}[h!]
\centering
\arrayrulecolor{black}
\caption{Pre/Post-survey, * --- Twitter emotional influence questions.}
\begin{tabular}{cccccccc}
& \multicolumn{1}{c}{(1)} & \multicolumn{1}{c}{(2)} & \multicolumn{1}{c}{(3)} & \multicolumn{1}{c}{(4)}  & \multicolumn{1}{c}{(5)} & \multicolumn{1}{c}{(6)} & \multicolumn{1}{c}{(7)} \\ \hline
\textbf{Pre}, * &   &   &   &   &   &   &   \\
Mean & \textbf{50.0} &  \textbf{50.8}  &  \textbf{56.4} &  \textbf{28.3} & 52.3 &  \textbf{52.7} &  \textbf{46.0} \\
Std & 27.1  & 28.2  &  26.8 &  23.5 &  29.5 &  30.7 &  23.1 \\\hline
\textbf{Post}, *  &   &   &   &   &   &   &   \\
Mean & \textbf{60.7}  & \textbf{60.6}  &  \textbf{69.4} &  \textbf{40.9} &  63.5 &  \textbf{70.8} &  \textbf{58.5} \\
Std & 24.3  & 24.5  &  14.3 &  16.9 &  25.5 &  23.0 &  20.1 \\\hline
P-value, * & 0.025 & 0.017 &  0.047 &  0.047 &  0.292 &  0.005 &  0.029 \\\hline\hline
\multicolumn{8}{l}{* participants who completed pre- and post-survey.}
\end{tabular}
\label{table:surveys-2}
\end{table}

We analyzed post-survey responses for \texttt{T1} and \texttt{T2} altogether. We expected an increase in T-Moodifier users' perception of emotional contagion (Questions (1)-(4)) while using T-Moodifier. As shown in Table \ref{table:surveys-2}, {\color{black}this perception changed significantly in the expected direction for all four questions}. 

{\color{black}
Regarding Questions (1) and (2), after the T-Moodifier experience, users could recognize significantly more strongly that Twitter and their connections influenced them, suggesting that they gained awareness thanks to T-Moodifier. As for changes in Question (3)}, they might indicate that users perceived more clearly what their friends shared and, hence, believed that what was happening on Twitter influenced their friends' reactions. In addition, changes in Question (4) might reflect that users updated their belief regarding their power to influence others.

{\color{black}Even though Question (5) did not change significantly, it moved in the expected direction towards increasing users' emotional reflection before posting content.}

On the subject of the users' confidence regarding the emotions they shared and consumed on Twitter, we observed the following in the responses to Questions (6) and (7): the awareness of T-Moodifier users increased and they could assert more confidently the main emotion they shared and perceived on Twitter. {\color{black} For the emotions asked for Question (6), users changed from 41\% in accuracy before the experience to 50\% in accuracy afterwards; for the emotions asked for Question (7), they changed from 42\% to 58\%. 

We conclude that these changes in awareness captured by the surveys, coupled with the observed behavioral changes, reflect T-Moodifier's potential to make users reflect more on what they see and share, increasing their emotional awareness.}\\

\subsubsection*{\textbf{Open-ended comments and design implications}}

Though we did not tell the participants that we would measure (publicly visible) changes in their behavior, in the open-ended comments, some of them acknowledged that they did reflect on how they reacted while using T-Moodifier. For example, one \texttt{T2} participant (who did not see their own statistics), stated ``[T-Moodifier] made me realize how I react to my feed's composition and people's positive or negative news.''
{\color{black}Furthermore, 75\% of those who returned the post-survey stated that T-Moodifier helped them better understand the emotions they consumed on Twitter. Most of them explicitly pointed to the tool's design as the means for that. They said that ``the filtration system [...] was a huge advantage,'' as well as ``the ability to toggle between positive and negative emotions.''  They also pointed to the labels and colored borders as a way of differentiating and ``giving more attention to emotions,'' or  by giving them ``a reason to hope they [Tweets] would be positive before I even read them.'' These are examples of how T-Moodifier provoked emotional reflection, recognition, and awareness among social media users.}

\subsubsection*{\textbf{Why did people decide to use T-Moodifier?}} 
Our pre-survey efforts included asking participants to comment on why they were installing T-Moodifier. One-third of them indicated that they decided to use T-Moodifier for various personal reasons. Another third of the participants indicated that they started using T-Moodifier because they were curious about the tool and how to use it. The final third stated that they were very interested in understanding and exploring their News Feed and how it affected them in greater depth.

Some people used the words ``unconsciously'' (ex: ``[I am] interested to see what I've followed unconsciously'') and ``subconsciously'' (ex: ``I want to understand what material I am subconsciously ingesting''). This indicated that they might have known they were unconsciously perceiving emotions on social media, but they did not have a tool, like T-Moodifier, to make those emotions explicit.

\subsection{Self-reported mental health}
{\color{black}
Previous works suggest that emotional awareness and recognition are crucial to wellbeing~\cite{schutte2011emotional, hill2012mindfulness, boden2015investigation} and can help to exploit emotions that improve one's mental health~\cite{lieberman2011subjective}. 

We designed T-Moodifier to help with these and hope that, under prolonged use, it could help improve Twitter users' emotional awareness and, consequently, their mental wellbeing. To approach this assessment, we applied the Eight-Item Personal Health Questionnaire (PHQ-8)~\cite{kroenke2009phq} as a preliminary way to understand whether tools like T-Moodifier could have an impact on users' mental health. Participants who completed both pre- and post-questionnaire responses scored around 5.6 in both cases on average, with 7/55 (12.7\%) of them scoring at least 10 (which is considered a sign of depressive disorders); there were no significant differences between the two treatments. We can draw only limited conclusions based on the 28 returned post-surveys; however, we can say that people in this target group are willing to use applications that are aimed at helping them. 

We do not have statistically significant evidence to support/refute our hypothesis regarding T-Moodifier helping users' wellbeing. For future studies, we plan to use the Warwick-Edinburgh Mental Wellbeing Scale~\cite{tennant2007warwick}, a more positively worded and less clinical-centric test, for assessment; this might also prevent concerns about inducing negative emotions instead of preventing them due to the application of a depression-centric test.}

\subsection{Users' inquiries and feedback about Tweet Moodifier}

Prior to running the experiment reported here, we ran a pilot version {\color{black}to get feedback on the T-Moodifier user experience}. Some participants in the pilot contacted us to ask for more details regarding the use of private information being collected, to know how T-Moodifier classifies Tweets' affective valences, and to report mislabeled Tweets. Consequently, we updated T-Moodifier and its website to provide all that information and implementation details from the beginning of the reported T-Moodifier experience. These questions show that transparency is key to reaching users and having them participate in this type of studies. As stated by Amodei et al~\cite{amodei2016concrete}, transparency is an emerging topic of concern among tech users.

\section{Conclusions}
\label{sec:conclusion}
We presented a new tool, Tweet Moodifier (T-Moodifier), a user-consented Chrome extension that aims to help users better understand the emotional valence of what they consume and share online when using Twitter. By allowing users to filter and make explicit the emotional content in their News Feed, T-Moodifier supports users in reflecting more confidently on the positive and negative nature of what emotions they consume and share. While the survey results in this work are preliminary (because of the small sample size completing both pre- and post-questionnaires), the behavioral results showed a significant association between receiving personalized statistical feedback on Tweet valence and an increase in the percentage of neutral Tweets sent. This result suggests that the use of T-Moodifier may help enhance Twitter users emotional awareness and may also influence their Twitter behavior. 

\section{Discussion and future work}
\label{sec:discussion}

The reported results have several limitations that should be addressed in the future to obtain stronger conclusions and to explore other aspects of Twitter users' emotional awareness that T-Moodifier currently does not capture.

While participation yielded 385 days of T-Moodifier use, a major limitation is that the current sample of returned post-surveys is only 28, and the total days of use per person is only 7. Thus, the conclusions based on self-report are not as strong as those based on behavior, and the behavior is relatively short-term. Our next step will entail shedding light on what the long-term effects of T-Moodifier on a bigger audience might be. This could allow us to generalize our conclusions and increase our understanding of how social media affects us and how aware we are of that. Also, with having a bigger audience, we could differentiate between clusters of users, such as strongly negative-/positive-sharing users, or even focus the study on groups with depressive disorders or social media dependency. Understanding diverse users' awareness might be a significant step toward influencing them positively.

Another major limitation is that this study reduced the emotions of Tweets to positive and negative valences. This reduction introduces noise mainly to negative emotions, where we can find hate speech and compassionate expressions of grief under the same \textit{negative} label. It has been shown~\cite{fan2014anger} that different emotions produce different reactions in social media users. Hence, in further versions, T-Moodifier should be able to break down both positive and negative valences into more specific emotions. Also, we want to explore more sophisticated models to improve the emotions classifier accuracy.

While T-Moodifier could potentially reach all Twitter users, an important limitation is that its current version is only available for Google Chrome in a desktop version for English Tweets. It is shown that most users access social media using mobile devices~\cite{han2017we}, and that people's sharing patterns are different across devices. Therefore, it is highly probable that T-Moodifier will not capture all representative patterns. We would also like to understand the awareness of those users who are not willing to use applications like T-Moodifier.

Furthermore, by using T-Moodifier's current capabilities, we could try to understand what would be a healthy emotional diet/regimen for people in social media. So far, we explored participants organic behavior once they are aware of the emotional content that they perceive. We plan on extending this to give users the option to balance how much (percentage-wise) of each emotion they want to receive \cite{hunt2018no,primack2017social}. 

{\color{black}As for T-Moodifier's design, we observed that the display of explicit feedback (i.e., personal statistics) revealing the valence of what users posted impacted them significantly. Also, users acknowledged that making emotions salient via colored marks and emojis in each Tweet helped them to differentiate and recognize emotions throughout their exposure to Twitter better. For a future version, we plan on giving a more positive tone to the availability of mental health resources by framing them as wellbeing resources, which is what T-Moodifier aims to be.}

Overall, we can see that T-Moodifier appears to be able to increase users' emotional awareness. However, it is too soon to say whether T-Moodifier causes a positive, negative or neutral effect in its users. We believe that prolonged use of tools that subtly elicit user emotional awareness could reduce the negative consequences of spending time in social media and help users take better control over their affective well-being.

\section*{Acknowledgment}
\footnotesize{}
The authors would like to thank the Network Computing Systems at the MIT Media Lab for providing the hardware required for deploying Tweet Moodifier. This research project was made possible thanks to the continuous guidance of the Committee on the Use of Humans as Experimental Subjects.

\bibliography{b}{}

\begin{thebibliography}{10}

\bibitem{ahn2013show}
June Ahn and Jinyoung Kim.
\newblock The show must go on: the presentation of self during interpersonal
  conflict on facebook.
\newblock In {\em Proceedings of the 76th ASIS\&T Annual Meeting: Beyond the
  Cloud: Rethinking Information Boundaries}, page~94. American Society for
  Information Science, 2013.

\bibitem{amodei2016concrete}
Dario Amodei, Chris Olah, Jacob Steinhardt, Paul Christiano, John Schulman, and
  Dan Man{\'e}.
\newblock Concrete problems in ai safety.
\newblock {\em arXiv preprint arXiv:1606.06565}, 2016.

\bibitem{barrett2001knowing}
Lisa~Feldman Barrett, James Gross, Tamlin~Conner Christensen, and Michael
  Benvenuto.
\newblock Knowing what you're feeling and knowing what to do about it: Mapping
  the relation between emotion differentiation and emotion regulation.
\newblock {\em Cognition \& Emotion}, 15(6):713--724, 2001.

\bibitem{boden2011you}
Matthew~Tyler Boden and Howard Berenbaum.
\newblock What you are feeling and why: Two distinct types of emotional
  clarity.
\newblock {\em Personality and individual differences}, 51(5):652--656, 2011.

\bibitem{boden2015investigation}
Matthew~Tyler Boden, Jessica~G Irons, Matthew~T Feldner, Sarah Bujarski, and
  Marcel~O Bonn-Miller.
\newblock An investigation of relations among quality of life and individual
  facets of emotional awareness and mindfulness.
\newblock {\em Mindfulness}, 6(4):700--707, 2015.

\bibitem{coviello2014detecting}
Lorenzo Coviello, Yunkyu Sohn, Adam~DI Kramer, Cameron Marlow, Massimo
  Franceschetti, Nicholas~A Christakis, and James~H Fowler.
\newblock Detecting emotional contagion in massive social networks.
\newblock {\em PloS one}, 9(3):e90315, 2014.

\bibitem{erisman2010preliminary}
Shannon~M Erisman and Lizabeth Roemer.
\newblock A preliminary investigation of the effects of experimentally induced
  mindfulness on emotional responding to film clips.
\newblock {\em Emotion}, 10(1):72, 2010.

\bibitem{fan2014anger}
Rui Fan, Jichang Zhao, Yan Chen, and Ke~Xu.
\newblock Anger is more influential than joy: Sentiment correlation in weibo.
\newblock {\em PloS one}, 9(10):e110184, 2014.

\bibitem{ferrara2015measuring}
Emilio Ferrara and Zeyao Yang.
\newblock Measuring emotional contagion in social media.
\newblock {\em PloS one}, 10(11):e0142390, 2015.

\bibitem{giachanou2016like}
Anastasia Giachanou and Fabio Crestani.
\newblock Like it or not: A survey of twitter sentiment analysis methods.
\newblock {\em ACM Computing Surveys (CSUR)}, 49(2):28, 2016.

\bibitem{go2009twitter}
Alec Go, Richa Bhayani, and Lei Huang.
\newblock Twitter sentiment classification using distant supervision.
\newblock {\em CS224N Project Report, Stanford}, 1(12), 2009.

\bibitem{grosser2014metrics}
Benjamin Grosser.
\newblock What do metrics want? how quantification prescribes social
  interaction on facebook.
\newblock {\em Computational Culture}, 2014.

\bibitem{han2017we}
Youngsub Han, Beomseok Hong, Hyeoncheol Lee, and Kwangmi Kim.
\newblock How do we tweet? the comparative analysis of twitter usage by message
  types, devices, and sources.
\newblock {\em The Journal of Social Media in Society}, 6(1):189--219, 2017.

\bibitem{hill2012mindfulness}
Christina~LM Hill and John~A Updegraff.
\newblock Mindfulness and its relationship to emotional regulation.
\newblock {\em Emotion}, 12(1):81, 2012.

\bibitem{huang2013distinguishing}
Shanshan Huang, Howard Berenbaum, and Philip~I Chow.
\newblock Distinguishing voluntary from involuntary attention to emotion.
\newblock {\em Personality and Individual Differences}, 54(8):894--898, 2013.

\bibitem{hunt2018no}
Melissa~G Hunt, Rachel Marx, Courtney Lipson, and Jordyn Young.
\newblock No more fomo: Limiting social media decreases loneliness and
  depression.
\newblock {\em Journal of Social and Clinical Psychology}, pages 751--768,
  2018.

\bibitem{kramer2014experimental}
Adam~DI Kramer, Jamie~E Guillory, and Jeffrey~T Hancock.
\newblock Experimental evidence of massive-scale emotional contagion through
  social networks.
\newblock {\em Proceedings of the National Academy of Sciences}, page
  201320040, 2014.

\bibitem{kroenke2009phq}
Kurt Kroenke, Tara~W Strine, Robert~L Spitzer, Janet~BW Williams, Joyce~T
  Berry, and Ali~H Mokdad.
\newblock The phq-8 as a measure of current depression in the general
  population.
\newblock {\em Journal of affective disorders}, 114(1-3):163--173, 2009.

\bibitem{lieberman2011subjective}
Matthew~D Lieberman, Tristen~K Inagaki, Golnaz Tabibnia, and Molly~J Crockett.
\newblock Subjective responses to emotional stimuli during labeling,
  reappraisal, and distraction.
\newblock {\em Emotion}, 11(3):468, 2011.

\bibitem{panger2016reassessing}
Galen Panger.
\newblock Reassessing the facebook experiment: critical thinking about the
  validity of big data research.
\newblock {\em Information, Communication \& Society}, 19(8):1108--1126, 2016.

\bibitem{primack2017social}
Brian~A Primack, Ariel Shensa, Jaime~E Sidani, Erin~O Whaite, Liu yi~Lin,
  Daniel Rosen, Jason~B Colditz, Ana Radovic, and Elizabeth Miller.
\newblock Social media use and perceived social isolation among young adults in
  the us.
\newblock {\em American journal of preventive medicine}, 53(1):1--8, 2017.

\bibitem{rainie2012tone}
Lee Rainie, Amanda Lenhart, and Aaron Smith.
\newblock The tone of life on social networking sites.
\newblock {\em Pew Internet Report}, 2012.

\bibitem{rosa2019automatic}
Hugo Rosa, N~Pereira, Ricardo Ribeiro, Paula~Costa Ferreira, Jo{\~a}o~Paulo
  Carvalho, S~Oliveira, Lu{\'\i}sa Coheur, Paula Paulino, AM~Veiga Sim{\~a}o,
  and Isabel Trancoso.
\newblock Automatic cyberbullying detection: A systematic review.
\newblock {\em Computers in Human Behavior}, 93:333--345, 2019.

\bibitem{schmidt2017survey}
Anna Schmidt and Michael Wiegand.
\newblock A survey on hate speech detection using natural language processing.
\newblock In {\em Proceedings of the Fifth International Workshop on Natural
  Language Processing for Social Media}, pages 1--10, 2017.

\bibitem{schutte2011emotional}
Nicola~S Schutte and John~M Malouff.
\newblock Emotional intelligence mediates the relationship between mindfulness
  and subjective well-being.
\newblock {\em Personality and Individual Differences}, 50(7):1116--1119, 2011.

\bibitem{sleeper2013post}
Manya Sleeper, Rebecca Balebako, Sauvik Das, Amber~Lynn McConahy, Jason Wiese,
  and Lorrie~Faith Cranor.
\newblock The post that wasn't: exploring self-censorship on facebook.
\newblock In {\em Proceedings of the 2013 conference on Computer supported
  cooperative work}, pages 793--802. ACM, 2013.

\bibitem{tennant2007warwick}
Ruth Tennant, Louise Hiller, Ruth Fishwick, Stephen Platt, Stephen Joseph,
  Scott Weich, Jane Parkinson, Jenny Secker, and Sarah Stewart-Brown.
\newblock The warwick-edinburgh mental well-being scale (wemwbs): development
  and uk validation.
\newblock {\em Health and Quality of life Outcomes}, 5(1):63, 2007.

\bibitem{tromholt2016facebook}
Morten Tromholt.
\newblock The facebook experiment: Quitting facebook leads to higher levels of
  well-being.
\newblock {\em Cyberpsychology, behavior, and social networking},
  19(11):661--666, 2016.

\bibitem{doi:10.1111/j.1468-2958.2007.00312.x}
Joseph~B. Walther, Brandon Van Der~Heide, Sang-Yeon Kim, David Westerman, and
  Stephanie~Tom Tong.
\newblock The role of friends’ appearance and behavior on evaluations of
  individuals on facebook: Are we known by the company we keep?
\newblock {\em Human Communication Research}, 34, 2008.

\bibitem{wulczyn2017ex}
Ellery Wulczyn, Nithum Thain, and Lucas Dixon.
\newblock Ex machina: Personal attacks seen at scale.
\newblock In {\em Proceedings of the 26th International Conference on World
  Wide Web}, pages 1391--1399. International World Wide Web Conferences
  Steering Committee, 2017.

\bibitem{zhao2008identity}
Shanyang Zhao, Sherri Grasmuck, and Jason Martin.
\newblock Identity construction on facebook: Digital empowerment in anchored
  relationships.
\newblock {\em Computers in human behavior}, 24(5):1816--1836, 2008.

\end{thebibliography}
\bibliographystyle{plain}

\appendix
\label{sec:appendix}
\footnotesize{}
\vspace{-1.5em}

\begin{figure}[h!]
    \centering
    \includegraphics[width=.6\linewidth]{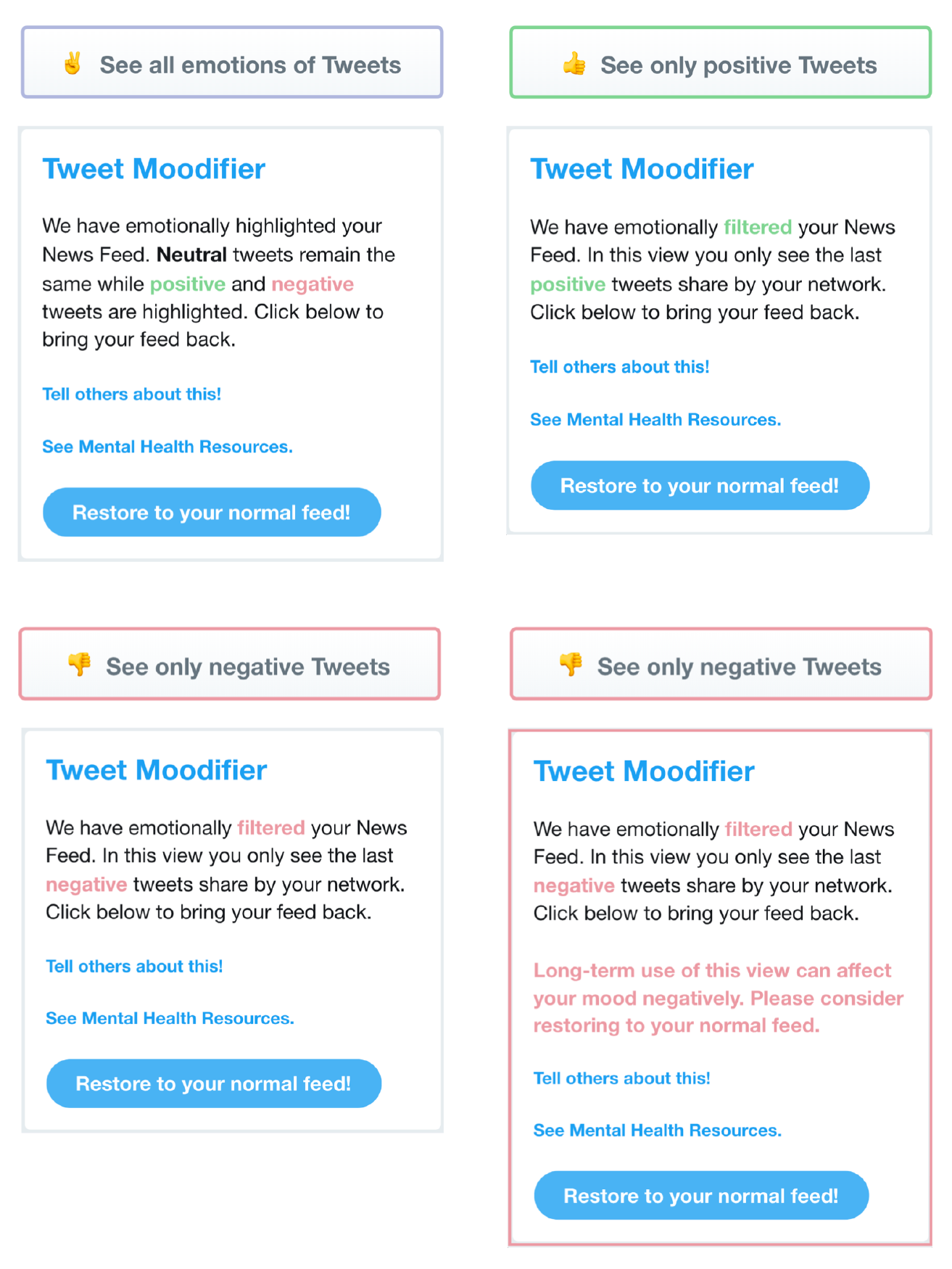}
    \vspace{-.5em}
    \caption{T-Moodifier updates its card to explain the activated view.}
    \label{fig:cards}
\end{figure}


\vspace{-1.5em}
\begin{figure}[h!]
    \centering
    \includegraphics[width=.55\linewidth]{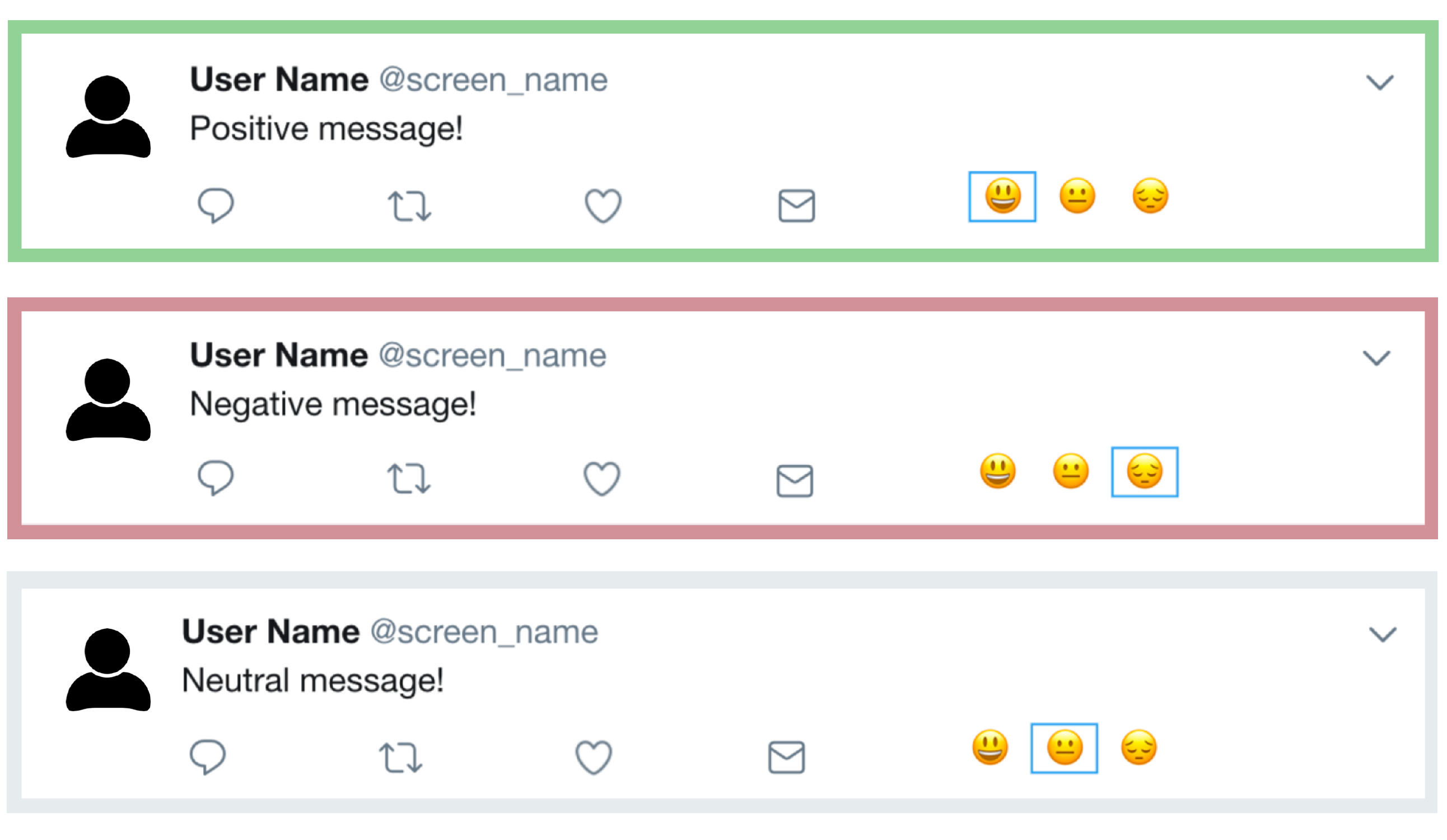}
    \vspace{-.5em}
    \caption{``Moodified'' Tweets.}
    \label{fig:tweets}
\end{figure}

\end{document}